\documentclass{PoS}

\title{Top quark measurements by the D0 collaboration}

\ShortTitle{D0 top quark}

\author{\speaker{Paul Grannis}, for the D0 Collaboration \\
        Stony Brook University\\
        E-mail: \email{pgrannis@sunysb.edu}}

\abstract{The 2019 European Physical Society Prize for High Energy and Particle
Physics was awarded to the CDF and D0 collaborations for "the discovery of
the top quark and the detailed measurement of its properties".   This paper 
is based upon the talk accepting the prize on behalf of the D0 Collaboration.
}

\FullConference{
European Physical Society Conference on High Energy Physics - EPS-HEP2019 -\\
			10-17 July, 2019\\
			Ghent, Belgium}

\begin{document}

In 1981, Fermilab Director Leon Lederman issued a call for an experiment to 
be located in the D0 interaction region for "a small (9 m cube), simple 
and clever" experiment.  About 20 expressions of interest were narrowed to 
12 proposals that were then considered by the  Physics Advisory Committee in 
summer 1983.  All proposals were rejected in favor of a new, completely 
unspecified experiment, charged to build a detector that would be
"no worse than those proposed."   The collaboration that then formed prepared a 
design~\cite{run1_detector} that emphasized hermetic, finely segmented, 
uranium liquid-argon calorimetry and a large-coverage muon detection 
system with stations before and after solid iron magnetized toroids.  The
tracking detectors, including a transition radiation detector, were placed within 
the 75 cm inner radius of the calorimetry in a region with no magnetic field.
It was deemed that it would be too expensive to make a bypass for the 
400 GeV Main Ring accelerator, used to feed the 
fixed target experiments and to create the antiprotons for the Tevatron, 
as had been done
at CDF. So it was left to penetrate the D0 liquid argon calorimeter, thus
requiring that data taking be suspended when particles were in the
Main Ring.

\begin{figure}
\includegraphics[width=1.0\textwidth]{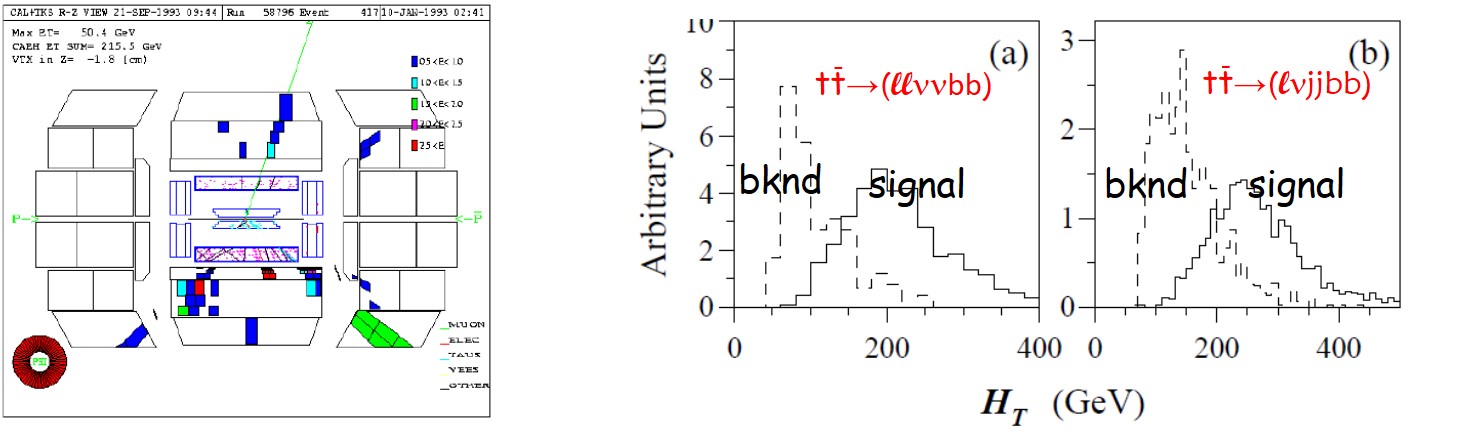}
\caption{(left) Event display of an event with an electron ($p_T=100$ GeV), muon
(200 GeV), missing transverse momentum (100 GeV) and two jets of 25 and 20 GeV.
(right) The $H_T$ distributions expected for top pair signal and backgrounds for 
(a) dilepton  and (b) single lepton events.
}
\label{event417}
\end{figure}

D0 started its Run I operation in 1992, five years after CDF had its first run (CDF's 
head start was mitigated by the rapid increase in the Tevatron luminosity).
The second D0 physics publication in 1994~\cite{toplimit_1994} set the last-ever lower 
 limit on the mass of the top quark at 131 GeV.  For these masses 
top decays to $W b$ 100\% of the time, so that $t\overline t$ events
are categorized according to the decays of the two $W$ bosons (to leptons or quarks), resulting in
$t\overline t$ classifications of dilepton, single lepton, or all jets.
That paper presented a 
striking event (Fig.~\ref{event417}) with a high $p_T$ electron and muon, missing 
transverse momentum
 and two jets, for which the 
background was very small. 
The most  likely top mass from this single event was 145 GeV.  

In 1994, CDF published evidence for the top quark with a mass near
175 GeV.  At the time, D0 had similar sensitivity but fewer events.  As a result of these
indications, D0 updated its analysis strategy to focus on the higher mass region.  In the 
 absence of a magnetic field and the ability to tag $b$-quarks by their finite lifetime,
D0 relied upon the topological variables $H_T$, the scalar sum of the transverse 
momenta of jets and leptons (see Fig.~\ref{event417}),
and Aplanarity (the tendency for the object momenta
to be distributed uniformly in 3-space).

After realigning some magnets in summer 1994, the Tevatron instanteous luminosity
grew rapidly so that by early 1995 it became apparent that there was now enough data  
to make a discovery.   CDF and D0 had previously reached an agreement with Director 
John Peoples that when one of the collaborations announced that they were ready to 
submit for publication, the other collaboration would have one week to finalize its
analysis for a joint submission.   On Feb. 18, CDF started the clock.  D0 had a manuscript
in hand but chose to spend a bit more time to check
the systematic uncertainties and significance calculations.  
The papers ~\cite{discovery} were simultaneously 
submitted to Phys. Rev. Letters on Feb. 25.  

\begin{figure}
\includegraphics[width=0.45\textwidth]{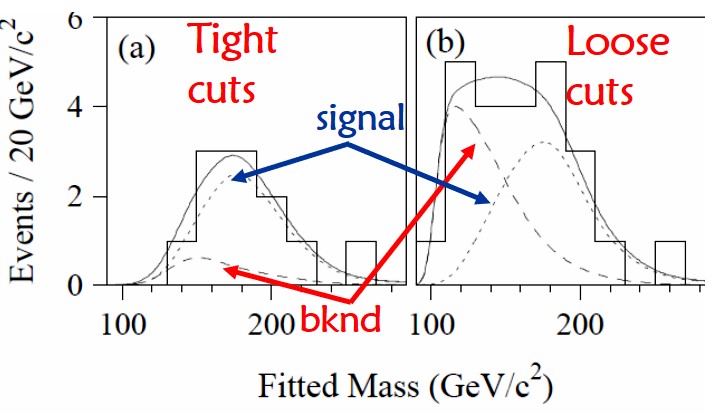}
\includegraphics[width=0.05\textwidth]{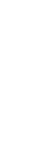}
\includegraphics[width=0.45\textwidth]{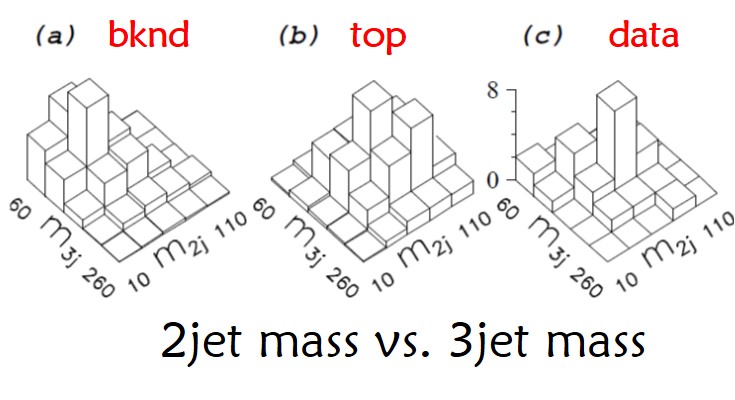}
\caption{ 
 (left) Number of events vs. fitted top quark mass for the single lepton selection, for both tight and loose selection cuts.  (right) Two jet mass vs. three jet mass for the top quark decaying to three jets in the single lepton selection.   The data shows a clear contribution from the $t \rightarrow W + b$-jet contribution. 
}
\label{prl_observation}
\end{figure}

\begin{figure}
\includegraphics[width=0.9\textwidth]{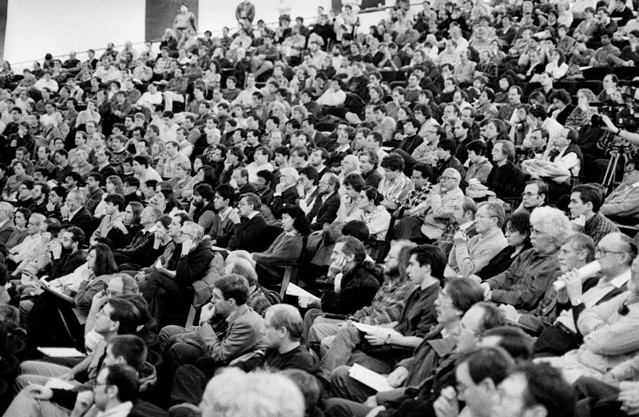}
\caption{ 
 The audience at the Mar. 2, 1995 top discovery seminars.   We don't know whether the speaker was CDF or D0, but note that no one is glued to their phone or laptop! 
}
\label{announcement}
\end{figure}

The D0 analysis, based on 50 pb$^{-1}$ of $p\overline p$ collision data at 1.8 TeV considered
the dilepton channels ($ee$, $e\mu$ and $\mu\mu$),  and single lepton channels ($e$ or $\mu$) with
and without tagging of $b$-jets by a decay muon.
The separation of signal and background 
for the D0 single lepton sample for both tight and loose selection cuts, and the two-dimensional 
histogram of the  three jet mass vs. two jet mass for background, expected top pair signal and data are
 shown in Fig.~\ref{prl_observation}. An embargo was placed on the results until the public 
announcement on Mar. 2 (Fig.~\ref{announcement}), delayed since several of the principals were
at a conference in Brazil reporting on the now-obsolete evidence results.  
Nevertheless several newspapers picked up the rumors of an impending discovery and attempted
scoops.

Run 1 continued until 1996 with a total accumulation of about 100 pb$^{-1}$ and led to 16 
top quark publications
that focused mainly on measurements of the $t\overline t$ cross section 
 and top quark mass.  
These papers also included the first measurement of the $t$ and $\overline t$ spin
correlation and searches for the electroweak single top production process and for $t\overline t$ resonances.

The Tevatron had a long shutdown from 1996 to 2001 to complete the Main Injector, remove
the Main Ring, increase the collision energy to 1.96 TeV, install helical orbits to separate $p$
and $\overline p$, and other improvements.   During this shutdown major upgrades
were completed for the D0 detector~\cite{run2upgrade}.   
A 2T solenoid magnet was introduced just inside the central
calorimeter.  The tracking system was completely rebuilt with a four (later five) layer silicon 
strip barrel and disk detector and an eight layer scintillating fiber tracker detector read out by
precursors of solid state photomultipliers.  The calorimetry was extended to include
preshower detectors on the front faces of the cryostats and the 
shaping electronics were replaced to allow operation with bunch spacing interval 
reduction from 3 $\mu$s to 396 ns.   New scintillator pixel panels were installed before and
after the muon toroids to improve muon triggering, and new small cell drift tube position
detectors were installed in the forward regions.  Triggering was improved with new track
and preshower-based triggers and a new level of triggering was introduced.   The software was
rewritten entirely in C++.  The upgraded detector was capable of collecting data at rates up to 
1000 Hz.   These improvements and the 100-fold increase in integrated luminosity during Run II 
enabled large improvements in physics reach and precision for top quark studies.  

\begin{figure}
\includegraphics[width=0.55\textwidth]{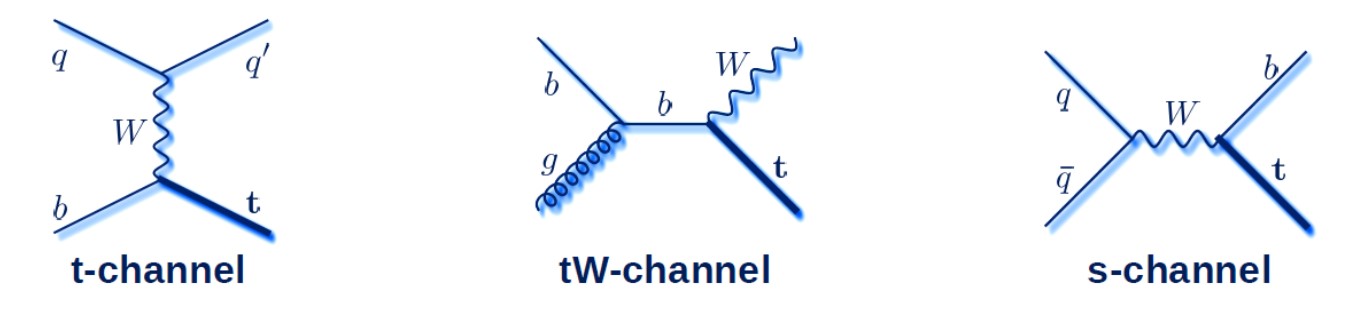}
\includegraphics[width=0.05\textwidth]{space.jpg}
\includegraphics[width=0.30\textwidth]{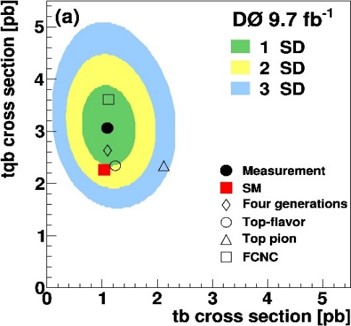}
\caption{ 
 Three diagrams contribute to single top quark production through the electroweak interaction.   Only the s-channel and t-channel processes have been observed at the Tevatron.  (The LHC experiments observe only the t-channel and $tW$ channel processes.)   As shown in the right panel, the comparison of the D0 s- and t-channel cross sections constrains models of new physics.
}
\label{singletop}
\end{figure}

\begin{figure}
\includegraphics[width=0.45\textwidth]{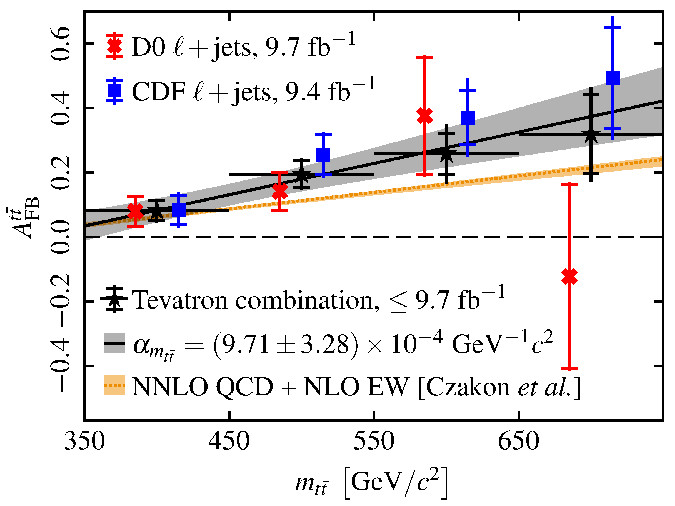}
\includegraphics[width=0.05\textwidth]{space.jpg}
\includegraphics[width=0.38\textwidth]{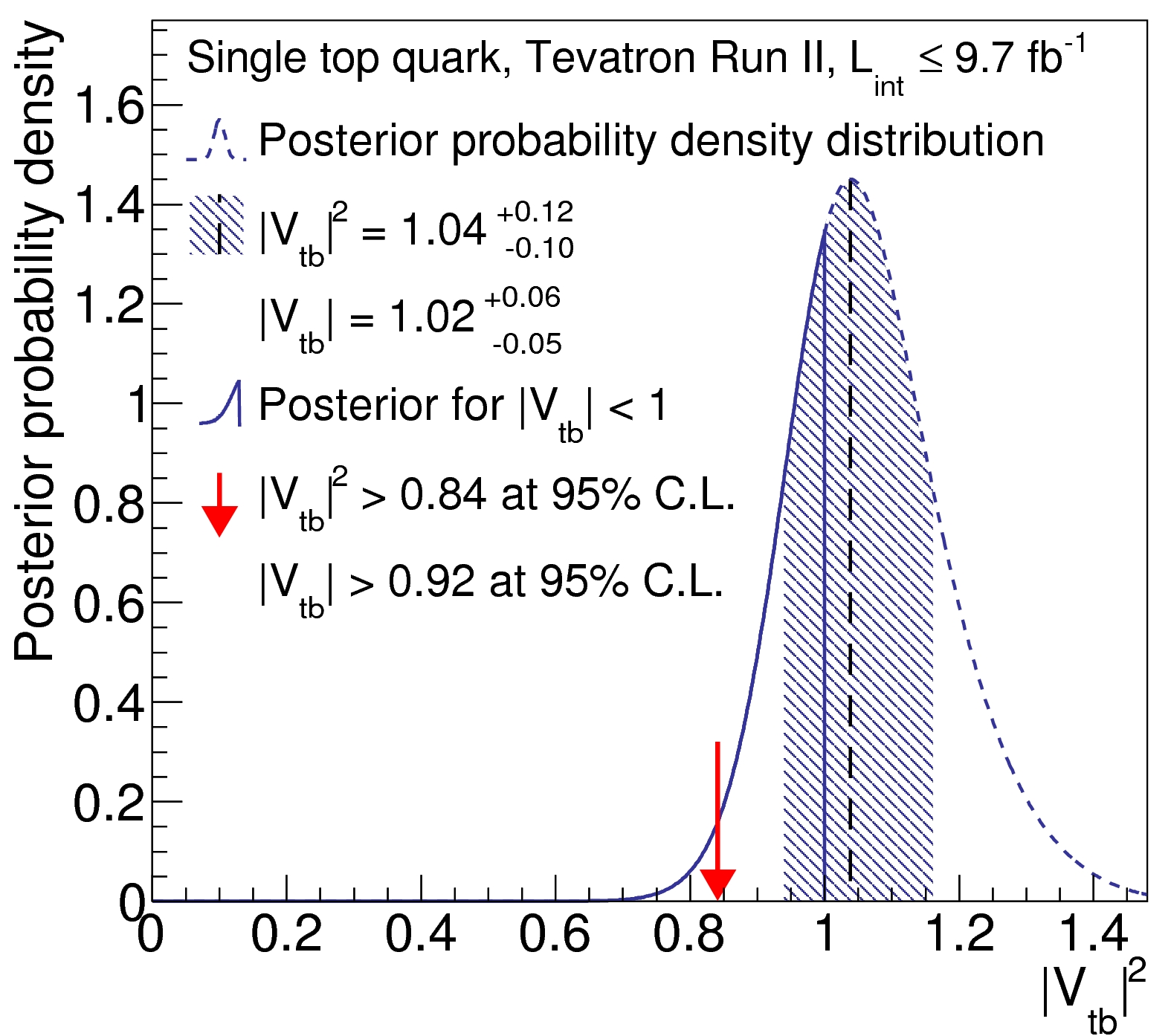}
\caption{ 
 (left) The combined CDF and D0 measurement of the forward-backward asymmetry 
for $t\overline t$ production as a function of the
mass of the $t\overline t$ system.   "Forward" ("backward")
corresponds to the top quark (antitop) being emitted in the hemisphere containing the outgoing beam 
protons.  
(right) The combined CDF and D0 measurement of the posterior probability density for the 
CKM parameter $|V_{tb}|^2$ .
}
\label{Afb}
\end{figure}

\begin{figure}
\includegraphics[width=0.52\textwidth]{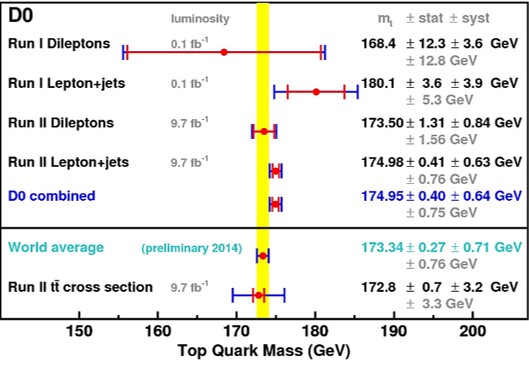}
\includegraphics[width=0.05\textwidth]{space.jpg}
\includegraphics[width=0.37\textwidth]{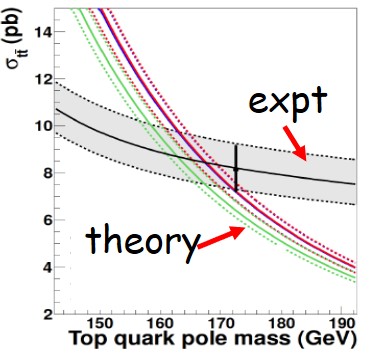}
\caption{ 
 (left) The D0 top quark mass measurements using the Monte Carlo template method.  The connection of 
this mass to theoretically well-defined masses is not precisely known.  (right) A preliminary
measurement~\cite{diff-pole-mass} of the top quark pole mass, $m_{\rm pole}=169.1\pm2.5$ GeV, 
based on the comparison 
of theoretical and measured differential 
$t\overline t$  cross sections as a function of $p_T^t$ and $M(t\overline t)$.
}
\label{topmass}
\end{figure}

There have been
91 top quark publications~\cite{d0-pubs} to date based on the D0  Run  2 data set.  There is 
insufficient space
in this report to discuss these results in detail, but a recent review can be found in
Ref.~\cite{uspekhi}.   The topics investigated include:

\begin{itemize}
\item $t\overline t$ differential and inclusive cross section measurements in dilepton, single lepton 
(including $\tau$+jets)  and all jets channels;
\item observation of single top quark production via the electroweak interaction;
\item observation of the distinct s- and t-channel single top production processes
(see Fig.~\ref{singletop});
\item observation of spin correlations in $t\overline t$  production;
\item measurement of top quark polarization in dilepton and lepton+jets production;
\item forward-backward asymmetries in top quark production and for the decay leptons 
(see Fig.~\ref{Afb});
\item search for Lorentz non-invariance in $t\overline t$ production and decay;

\vspace{3mm}
\item top quark mass measuremens using Monte Carlo templates (see Fig.~\ref{topmass});
\item top quark pole mass measurements using inclusive and differential $t\overline t$ cross sections
(see Fig.~\ref{topmass});
\item differentiation between  the standard model value of 2/3e for the top quark charge and 4/3e;
\item measurement of the width of the top quark state;
\item comparison of the mass of $t$ and $\overline t$ and test of CPT;
\item measurement of the helicity fractions of $W$ bosons in top decay;
\item measurements of anomalous top quark tensor and  axial vector couplings;
\item measurement of the ratio ($B(t\rightarrow Wb$))/($B(t\rightarrow W \Sigma q_i$))
\item measurement of the CKM parameter $V_{tb}$ (see Fig.~\ref{Afb});
\item study of color flow in $t\overline t$ events;
\vspace{3mm}
\item search for fourth generation top quarks;
\item search for flavor changing neutral currents in top  decay;
\item search for charged Higgs bosons in top decay;
\item search for narrow $t\overline t$ resonances;
\item search for admixtures of spin 0 and spin 1/2 top quarks;
\item search for $W'$ bosons in $W' \rightarrow t\overline b$;
\item search for anomalous $Wtb$ couplings in single top production

\end{itemize}

Although the kinematic range of the differential cross sections and most of the search results suffer in 
comparison with the recent LHC measurments, many of the measurements of the properties are 
competitive with the LHC, and some measurements such as the  
$t\overline t$ forward-backward asymmetry and the 
measurement of the s-channel single top quark cross section have been unique Tevatron contributions.

In an enterprise as complex as building a modern collider detector, operating it with maximum
efficiency, managing the flood of data that it produces, and generating physics results on scores of 
topics, it is impossible to single out one individual for recognition.   Thus the award of the 2019 EPS 
prize to the collaboration is most appropriate.  It is also apt that the prize citation mentions not
only the discovery of the top quark but also the detailed measurements of its properties.   Those
measurements involved collaboration members working over the full time span of the experiment and 
those achievements are as important as the original discovery.   Over the lifetime of the D0 collaboration 
some   1500 physicists contributed to its success, and it is appropriate that all join in expressing their
appreciation for the prize award.


\newpage

{\bf Acknowledgements}  
The outstanding performance of the Tevatron was essential for the success of the D0 program.
We greatly appreciate the outstanding work by the staff of the Fermilab Accelerator Division and
appreciate the support of the D0 physics program by the Fermilab Computing and Particle Physics 
divisions.
We thank the staffs at collaborating institutions,
and acknowledge support from the
DOE and NSF (USA);
CEA and CNRS/IN2P3 (France);
MON, NRC KI, and RFBR (Russia);
CNPq and FAPERJ (Brazil);
DAE and DST (India);
Colciencias (Colombia);
CONACyT (Mexico);
NRF (Korea);
FOM (The Netherlands);
STFC and The Royal Society (UK);
MSMT (Czech Republic); 
BMBF and DFG (Germany);
SFI (Ireland);
Swedish Research Council (Sweden);
CAS and CNSF (China);
and
MESU (Ukraine).


\begin{thebibliography}{99}
\bibitem{run1_detector}
S. Abachi {\sl et al.} "{\it The D0 Detector}", Nucl. Instrum Methods in Phys. Res. Sect. 
{\bf A}338, 185 (1994).

\bibitem{toplimit_1994}
S. Abachi {\sl et al.} "{\it Search for the Top Quark in $p\overline p$ Collisions at $\sqrt s = 1.8$ TeV}", Phys. Rev. Lett. {\bf 72}, 2138 (1994).

\bibitem{discovery}
F. Abe {\sl et al.} "{\it Observation of Top Quark Production in $p\overline p$ Collisions 
with the Collider Detector at Fermilab}", Phys. Rev. Lett. {\bf 74}, 2626 (1995);
S. Abachi {\sl et al.} "{\it Observation of the Top Quark}", Phys. Rev. Lett. {\bf 74}, 2632 
(1995).

\bibitem{run2upgrade}
V.M. Abazov  {\sl et al.} "{\it The Upgraded D0 Detector}",  
Nucl. Instrum. Methods in Phys. Res. Sect. {\bf A 565}, 463 (2006).

\bibitem{d0-pubs}
The full list of D0 Run 2 publications can be found at
https://www-d0.fnal.gov/d0\_publications/d0\_pubs\_list\_bytopic.html

\bibitem{uspekhi}
Eduard Boos {\sl et al.} "{\it The top quark (20 years after the discovery)}",
Physics Uspekhi {\bf 58} (12), 1133 (2015); {\tt arXiv 1509.03325 [hep-ex] (2015)}.

\bibitem{diff-pole-mass}
https://www-d0.fnal.gov/Run2Physics/WWW/results/top.htm


\end{thebibliography}
\end{document}